\documentstyle[aps,prl,epsf,twocolumn,floats]{revtex}

\begin{document}
\draft
\title{Current induced switching of magnetic domains 
to a perpendicular configuration}

\author{ X. Waintal and P. W. Brouwer}
\address{Laboratory of Atomic and Solid State Physics,
Cornell University, Ithaca, NY 14853-2501\\
{\rm \today} 
\medskip \\ \parbox{14cm}{\rm
In a ferromagnet--normal-metal--ferromagnet trilayer,
a current flowing 
perpendicularly to the layers creates a torque on the magnetic
moments of the ferromagnets. 
When one of the contacts is superconducting,
the torque not only favors parallel or antiparallel
alignment of the magnetic moments, as is the case for two normal
contacts, but can also favor a configuration where the two moments are
perpendicular. In addition, whereas
the conductance for parallel and antiparallel magnetic moments is the same,
signalling the absence of giant magnetoresistance in the usual sense, 
the conductance is greater in the perpendicular
configuration. Thus, a negative magnetoconductance is
predicted, in contrast with the usual giant magnetoresistance.
}}

\maketitle


A system made of stacks of alternating ferromagnetic and non-magnetic
metal layers shows Giant Magneto-Resistance (GMR) \cite{ibm}: When two
consecutive magnetic layers have their magnetic moments aligned, the
conductance is much bigger than when they are anti-aligned. A simple
way to think about this effect is in term of two separate
(i.e., incoherent)
currents for the majority and minority electrons (electrons with spin 
parallel or antiparallel to the magnetic moment,
respectively), and to
view the ferromagnetic layers as spin filters that have different
conductances for majority and minority electrons \cite{valet}.
In the configuration where the magnetic moments are aligned,
the majority electrons are well transmitted by both magnetic layers while 
the minority electrons are (mostly) reflected. When the moments are
anti-aligned, the majority spin direction of one layer is the minority 
spin direction of its neighbor so that all the electrons are reflected. The
role of the magnetic field is to align the initially anti-aligned
magnetic moments, thus giving rise to an increase of the conductance. 

When the magnetic
moments of the layers make an angle $\theta$ different from $0$ or
$\pi$, the simple ``two-current'' picture does not hold anymore;
instead a description in terms of 
a {\em coherent} superposition of spin up and spin down is needed. 
While the component of the spin flux 
in the direction of the magnetic moment $\vec m$ of a 
ferromagnetic layer
is conserved when an electric current is passed through the layer
(since the fluxes of majority and minority spins are 
conserved individually), the spin flux 
perpendicular to $\vec m$
does not have to be conserved, as it depends on the coherence
between majority and minority electrons.
As first
pointed out by Slonczewski~\cite{slon1} and Berger~\cite{berger},
the consequence of such a change of the magnetic moment 
carried by the current is that the current exerts a torque 
on the moments of
the ferromagnets. This so-called ``spin-transfer'' torque vanishes 
at the two angles $\theta=0$ and $\theta=\pi$, where the two current
model applies. Experimentally, one studies a
ferromagnet--normal-metal--ferromagnet (FNF) trilayer where one of the 
magnetic moments is held fixed, e.g.  by using a thick
ferromagnetic layer~\cite{myers,katine,wegrowe,grollier}, while the
other one is free to rotate. The
current direction determines which of the
two angles $\theta=0$ and $\theta=\pi$ is stable; switching the current
direction reverses the relative orientation of the magnets, and
thus changes the conductance. This signature of the current-induced
spin-transfer torque, which was observed experimentally in Refs
\cite{myers,katine}, 
provides a mechanism for a current controlled magnetic memory 
element. The spin-transfer torque can also be used for
other applications ~\cite{slon2}, 
including the excitation of spin-waves~\cite{tsoi}.

%

\begin{figure}
\epsfxsize=0.89\hsize
\hspace{0.05\hsize}
\epsffile{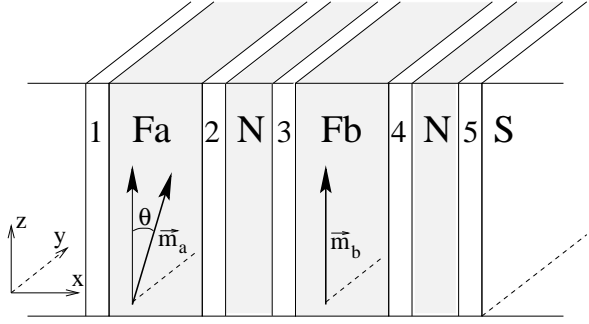}
\vglue +0.2cm
\caption{\label{system}
Schematic drawing of the system. Normal metal layers are $N_1$, $N_2$,
 and $N_3$, the ferromagnetic layers with moments $\vec m_a$ and
$\vec m_b$ in the $y$-$z$ plane 
are denoted with $F_a$ and $F_b$, and the superconductor
is denoted with $S$. The layers $N$ are metallic spacers, and
fictitious ideal leads $1$, $2$, $3$, $4$ and $5$ are added for
technical convenience.}
\end{figure}

If one of the two external contacts of an FNF trilayer is replaced
by a superconductor (S), see Fig.\ \ref{system},
the picture changes drastically. For voltages 
smaller than the superconducting gap, transport through the system 
occurs by Andreev reflection: an electron impinging on the 
S interface is reflected as a hole with opposite spin, adding a Cooper 
pair to the superconducting condensate. Each layer is therefore
traversed twice, once by a majority electron (or hole) and once
by a minority hole (or electron). 
Hence,  
the $\theta=0$ and
$\theta=\pi$ configurations have the same conductance, the difference
between them only being the order in which the spin filtering 
occurs
\cite{lambert}. 

What happens for angles $\theta$ other than $0$ or $\pi$? This 
question has a remarkable answer.
While the presence of Andreev reflection
suppresses the ``usual'' GMR at $\theta=0$ and $\pi$ (see previous
paragraph), 
we have found that it leads to 
a richer variety of effects for other $\theta$, where quantum 
coherence between spin up and spin down is crucial. With the S
contact, not only
parallel or antiparallel alignment of the magnetic moments plays a
special role, but also the perpendicular configuration:
(1) As with normal contacts, passing a current through 
the FNFS system exerts a torque on the moments of
the F layers; however, the possible stable 
configurations are not only $\theta=0$ and $\pi$, but also 
$\theta=\pi/2$ and $3\pi/2$. 
(2) The conductance $g$ is a $\pi$-periodic function of $\theta$ with
a maximum for $\theta=\pi/2$. 
These two results were derived for the experimentally relevant
case when
the ferromagnetic moment $\vec m_b$
neighboring the superconductor is held 
fixed, while the other moment $\vec m_a$
is allowed to vary, see Fig.\
\ref{system}, and assuming that the spin-relaxation length
is larger than the system size, which is a reasonable assumption
for a thin trilayer system.
By switching the current direction, one can switch the orientation
of the free magnetic moment $\vec m_a$ from
parallel to perpendicular to $\vec m_b$. Since $g(0) \neq g(\pi/2)$,
such a switch can be observed through a change in the conductance,
similar to what was observed in Refs.\ \cite{myers,katine} for the
case of normal contacts. A magnetic field will align the moments,
bringing the system from $\theta=\pi/2$ to $\theta=0$ (instead of
$\theta=\pi$ to $\theta=0$), so that one should observe a positive
magnetoresistance, in contrast to the standard GMR. 

The FNF trilayer under consideration is
shown in Fig.\ \ref{system}. The current flows in the $x$-direction, 
perpendicularly to the layers. For technical reasons, we have
added thin ideal (non-magnetic) spacers labeled $1,\ldots,5$ between
the F and N layers.  
Denoting the spin current in layer $i$ by $\vec{J}_i$
($i=1,\ldots,5$),
the torques $\vec{\tau}_b$ and $\vec{\tau}_a$ on the magnetic moments
$\vec m_a$ and $\vec m_b$ are given by
\begin{equation} \label{eq:torque}
  \vec{\tau}_b = \vec{J}_2 - \vec{J}_4,\ \
  \vec{\tau}_a=\vec{J}_1-\vec{J}_2.
\end{equation} 
Equation (\ref{eq:torque}) expresses that, unlike the probability 
current, the spin current $\vec J$ is not conserved by the ferromagnetic 
layers. (The component of $\vec J$ in the direction of the magnetization
is conserved, though.)
The difference in spin currents is transfered to the F
layers, as a torque acting on their magnetic moments.

\begin{figure}
\epsfxsize=0.89\hsize
\hspace{0.05\hsize}
\epsffile{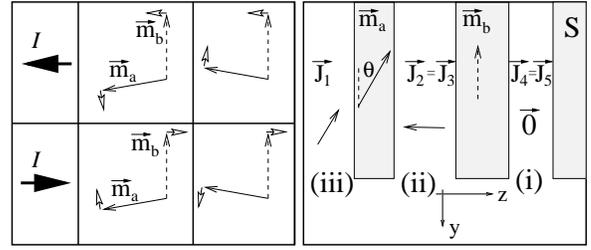}
\vglue +0.2cm
\caption{\label{signe} Left panel: signs of the torques (small arrows)
for different orientations of the moments $\vec{m}_a$ and $\vec{m}_b$
(long arrows). Upper (lower) panel: right (left) moving electrons. 
The torques lie in the plane spanned by $\vec{m}_a$ 
and $\vec{m}_b$. Right panel: schematic of the geometrical argument
leading to $\vec{\tau}_a=0$ in the perpendicular configuration.}
\end{figure}

We first give an intuitive explanation of how the presence of the
superconductor allows for the additional stable configurations at
$\theta=\pi/2$ and $\theta=3\pi/2$, and then present
the results of a more rigorous calculation. 
For the intuitive explanation 
we make the simplifying assumption 
that spins scattered from an F layer with moment $\vec m$
have their magnetic moment pointing parallel (or antiparallel) to
$\vec m$. (This assumption is valid if majority and minority
electrons are transmitted incoherently; It is not necessary for
the stability of $\theta=\pi/2$ or $3\pi/2$, as shown in our 
calculations below.)
With this assumption, the explanation proceeds as follows 
(see right panel of Fig~\ref{signe}). 
(i) 
For voltages below the
superconducting gap, only Cooper pairs can enter S. Hence, no spin
current can flow into S and one must have $\vec J_4 = \vec
J_5=0$. Therefore, by Eq.\ (\ref{eq:torque}), $\vec{\tau}_b =
\vec{J}_3$. 
(ii) 
Since the spin flux in the direction of $\vec m_b$ is 
conserved, there can be no component of the torque $\vec{\tau}_b$
directed along $\vec m_b$. Hence 
$\vec J_{2} = \vec J_{3}$ is perpendicular to $\vec m_b$. 
(iii)
Finally, for an unpolarized electron current entering the
trilayer, any spin current in layer $1$ must be carried by electrons
scattered from ${\rm F_a}$, so that $\vec{J}_1$ is parallel to 
$\vec m_a$. The torque  $\vec{\tau}_a$
is then given by the component of $\vec J_{2}$ perpendicular
to $\vec m_a$. Hence,
\begin{equation}\label{eq:intuit}
|\vec \tau_a| = |\vec \tau_b| | \cos \theta |
\end{equation} 
$\vec{\tau}_a$ thus vanishes at $\theta=\pi/2$ and $3\pi/2$
yielding the new possibility of stable configurations at these angles.


To make the above picture more precise, we use the
scattering approach \cite{beenakker1,jong}. 
The trilayer is bounded in the $y$ and $z$ directions, so 
that its transverse degrees of freedom are quantized, giving 
$N_{\rm ch}$ propagating modes (``channels'') at the Fermi level.
One has $N_{\rm ch} \sim A/\lambda_F^2$, $A$ being the cross section 
of the system and $\lambda_F$ the Fermi wave length.
Expanding the electronic wave function in these
modes, we can describe the system in terms of the $2 N_{\rm
ch}$-component vectors $\Psi_{i}^{e(h)L(R)}$, which is
the projection of the wave function onto the
left (right) going modes in region $i$ for the 
the electrons (holes) ($i=1,\ldots,5$). The $2 N_{\rm ch}$ components
of $\Psi$ account for the spin and channel degrees of
freedom. 
In the scattering approach, each layer is characterized by 
$2 N_{\rm ch} \times 2 N_{\rm ch}$ reflection
matrices $r_i$ and $r_i'$ and transmission matrices $t'_i$, $t_i$ 
(the label $i=1$ for ${\rm F_a}$, $i=3$ for ${\rm F_b}$, and $i=2,4$
for the two normal spacers),
\begin{eqnarray}
\Psi_{i+1}^{eR}&=& t_i \Psi_{i}^{eR} + r'_i \Psi_{i+1}^{eL},\ \
\Psi_{i+1}^{hR} = t_i^* \Psi_{i}^{hR} + {r'_i}^* \Psi_{i+1}^{hL},
\nonumber \\ 
\label{randt}
\Psi_{i}^{eL} &=&r_i \Psi_{i}^{eR} + t'_i \Psi_{i+1}^{eL}, \ \
\Psi_{i}^{hL}  = r_i^* \Psi_{i}^{hR} + {t'_i}^* \Psi_{i+1}^{hL}. 
\end{eqnarray}
At the NS interface electrons are reflected as holes,
\begin{equation}
  \Psi_{5}^{eL} = \sigma_y \Psi_{5}^{hR},\ \
  \Psi_{5}^{hL} = -\sigma_y \Psi_{5}^{eL}. \label{eq:andreev}
\end{equation}
To find the electrical and spin currents $I$ and $\vec J_i$ for
each layer $i$, we need
to calculate the 
generalized $2N_{\rm ch}\times 2N_{\rm ch}$ scattering matrices 
$S_{i,L(R)}^{e(h)e(h)}$ that
give the amplitudes of left (right) moving electrons (holes) in
region $i$ in terms of the amplitudes of the incoming
electron (hole) in the normal electrode (labeled with layer index
$i=1$),
\begin{equation}
\Psi_{i}^{e(h)L(R)}=S_{i,L(R)}^{e(h)e(h)} \Psi_{1}^{e(h)L}.
  \label{eq:sgen}
\end{equation}
Using Eq.(\ref{randt}), the matrices $S_{i,L(R)}^{e(h),e(h)}$
can be expressed in terms of $r_i$, $r'_i$, $t'_i$
and $t_i$. (This was done in Ref.\ \onlinecite{wmbr} for 
an FNF trilayer with two normal-metal contacts.)

Following the derivation of the Landauer
formula for the conductance \cite{StoneLesHouches}, one can now
calculate the electrical current $I$ \cite{LambertTakaneEbisawa} and 
the spin current $\vec J_i$ for an
applied voltage $V$ across the system as
%
\begin{eqnarray}\label{cond}
  {\partial I \over \partial V} &=& g = \frac{2e^2}{h} {\rm Tr}\, 
        S_{1,L}^{eh} S_{1,L}^{eh\dagger}, \\
  \label{spincond}
  \frac{\partial\vec{J_i}}{\partial V} &=& 
  -\frac{e}{4\pi}{\rm Re }\,
 {\rm Tr}\,  \left[\vec{\sigma}
S_{iR}^{ee}S_{iR}^{ee\dagger} \right.\nonumber \\ && \left. \mbox{}
 -\vec{\sigma}^* S_{iR}^{he}S_{iR}^{he\dagger} - \vec{\sigma}
S_{iL}^{ee}S_{iL}^{ee\dagger}
+\vec{\sigma}^* S_{iL}^{he}\Lambda_{iL}^{he\dagger}  
\right].
\end{eqnarray} 
where $\vec{\sigma}= (\sigma_x,\sigma_y,\sigma_z)$ is the vector of
Pauli matrices. 

It is important to notice that the scattering
matrices need to be calculated for each individual layer only
separately; solution of Eqs.\ (\ref{randt}) -- (\ref{eq:sgen}) then 
describes how
the information from the individual layers is combined to give 
properties of the multilayer system. 
For the ferromagnetic layers, the precise form of the reflection
and transmission matrices has to be determined from a microscopic
model, see e.g.\ Ref.\ \onlinecite{stiles}. For the normal metal
spacers, we make use of the 
polar decomposition \cite{beenakker1}, 
\begin{eqnarray}
\label{polar}
\left(
\begin{array}{cc} r_i & t'_i \\ t_i & r'_i  \end{array}
\right)= \openone_2 \otimes
\left(
\begin{array}{cc} u_i \sqrt{1-T_i} u'_i & i u_i \sqrt{T_i} v_i \\ 
  i v'_i \sqrt{T_i} u'_i & v'_i \sqrt{1-T_i} v_i \end{array} \right),
  \nonumber 
\end{eqnarray}
where $i=2,4$, $\openone_2$ is the $2 \times 2$ unit matrix in the spin 
grading, $u_i$ $u_i'$, $v_i$, and $v_i'$ are $N_{\rm ch} \times N_{\rm
ch}$ unitary matrices, and $T_i$ is a diagonal matrix containing the 
eigenvalues of $t_it_i^{\dagger}$ on the diagonal.
We consider an ensemble of trilayers for which the metallic spacers 
are good diffusive metals (with different impurity configurations for
different members of the ensemble), or for which 
the interfaces between the different material are rough, so that the
disorder mixes the different transverse modes in an isotropic way.
In that case one can describe the ensemble by taking the 
unitary matrices $u$, $u'$, $v$, and $v'$ uniformly distributed in 
the unitary group \cite{beenakker1}. We calculate an average over
the matrices $u$, $u'$, $v$, and $v'$ to find the ensemble average
of $\partial I/\partial V$ and $\partial \vec J_i/\partial V$.
Since, fluctuations are of relative size $1/N_{\rm ch}$,
 the average is sufficient to
characterize a single sample when $N_{\rm ch} \gg 1$.~\cite{beenakker1}
(In the experiments, typically $N_{\rm ch} \sim 10^3$.) 

After the ensemble average, the results
only depend on four parameters for each ferromagnetic
layer,~\cite{wmbr}
\begin{eqnarray}
  T_{i\uparrow} &=& N^{-1}_{\rm ch} \mbox{tr}\, t_{i\uparrow}
  t^{\dagger}_{i\uparrow}, \ \  
  T_{i\downarrow} = N^{-1}_{\rm ch} \mbox{tr}\, t_{i\downarrow}
  t^{\dagger}_{i\uparrow}, \nonumber \\
  T_{i\uparrow\downarrow} &=& N_{\rm ch}^{-1} \mbox{tr}\,
  t_{i\uparrow}t_{i\downarrow}^{\dagger},\ \
  R_{i\uparrow\downarrow}= N_{\rm ch}^{-1} \mbox{tr}\, r_{i\uparrow}
  r_{i\downarrow}^{\dagger}, \label{eq:mixing}
\end{eqnarray}
on the conductances of the normal spacers, and on the angle $\theta$.
In Eq.\ (\ref{eq:mixing}), the arrows refer to the majority and
minority spin directions.
The coefficients $T_{\uparrow\downarrow}$ and
$R_{\uparrow\downarrow}$ describe the coherence of transmission and/or
reflection of minority and majority spins. (The quantity
$N_{\rm ch} - N_{\rm ch}
R_{\uparrow\downarrow}$ is known as the ``mixing conductance'' 
\cite{brataas}.) The assumption
of ``incoherent'' transmission and reflection of majority and minority
that we made in the intuitive argument above, amounts to setting
$T_{\uparrow\downarrow}$ and $R_{\uparrow\downarrow}$ to zero. 
This is a fair assumption for 
$T_{\uparrow\downarrow}$, since majority and minority electrons
pick up different and uncorrelated phase shifts upon transmission 
through the ferromagnets, but it does not have to be the
case for $R_{\uparrow\downarrow}$ when the reflection at the 
interface is instantaneous. 
For ``perfect'' spin
filters, where all minority spins are reflected and all majority
spins are transmitted, $T_{\uparrow\downarrow}$ and
$R_{\uparrow\downarrow}$ are both zero. 

\begin{figure}
\epsfxsize=0.85\hsize
\hspace{0.0\hsize}
\epsffile{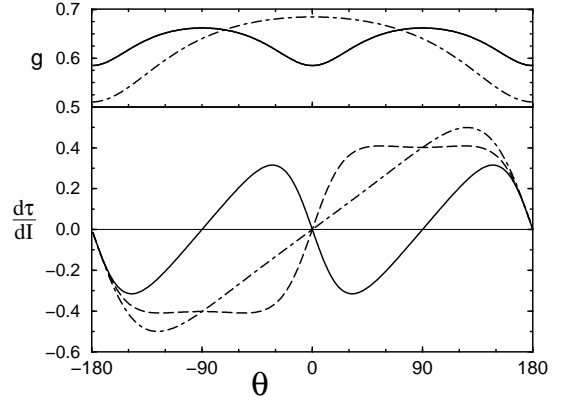}
\caption{\label{cocuco} Conductance (top) and torque per unit current 
(bottom) as a function of $\theta$ for a Co-Cu-Co trilayer system. 
$d\tau /dI$ is expressed in units of $h/4\pi e^2$, $g$ in units of
$N_{\rm ch} e^2/h$. We took $T_{a\uparrow}=T_{b\uparrow}=0.68$ and 
$T_{a\downarrow}=T_{b\downarrow}=0.29$. These numerical values were 
obtained from Ref.\ {\protect\cite{stiles}}. The conductance
of both normal spacers is taken equal to $2N_{\rm ch} e^2/h$,
corresponding to a situation where the main effect of disorder in
the spacers is
to randomize directions, not to backscatter electrons. 
The transmission/reflection coefficients $T_{\uparrow\downarrow}$ and
$R_{\uparrow\downarrow}$ have been set to zero. The solid (dashed)
curve shows $\tau_a$  ($\tau_b$) in the presence of the S contact. For
comparison, the dot-dashed curve shows $\tau_b=\tau_a$ for two normal 
contacts~\protect\cite{wmbr}.}
\end{figure}

The resulting 
expressions for the conductance and the torques are
rather lengthy, and will not be reported here. 
Nevertheless, they allow us to 
draw general conclusions about the direction and magnitude
of the torques for arbitrary values of the system parameters.
First, in agreement with the simple
argument given above, we find that the torques 
$\vec{\tau}_a$ and $\vec{\tau}_b$ lie in the 
plane spanned by the moments $\vec{m}_a$ and $\vec{m}_b$ [i.e.,
parallel to $(\vec{m}_a\times\vec{m}_b)\times\vec{m}_a$ and
$(\vec{m}_b\times\vec{m}_a)\times\vec{m}_b$ respectively].
Second, the magnitudes of the torques 
$\vec{\tau}_a$ and $\vec{\tau}_b$ are related as 
\begin{equation}\label{taub-a}
|\vec{\tau}_a| =- \frac{1- {\rm Re } \ T_{a\uparrow\downarrow}
- {\rm Re } \ R_{a\uparrow\downarrow}}{
1- {\rm Re } \ R_{a\uparrow\downarrow}}   |\vec{\tau}_b| |\cos\theta |,
\end{equation}
irrespective of the conductances of the normal layers. This equation
replaces Eq.(\ref{eq:intuit}) in case of non vanishing 
$R_{a\uparrow\downarrow}$ and 
$T_{a\uparrow\downarrow}$.
In Fig.~\ref{signe}, we have indicated the directions of 
$\vec \tau_a$ and $\tau_b$ for various relative orientations of
$\vec m_a$ and $\vec m_b$ and for the two possible directions of 
the current. 

In Fig.~\ref{cocuco}, the torques 
$\vec{\tau}_a$ and $\vec{\tau}_b$ and the conductance $g$ are 
shown versus $\theta$, for realistic
choices of the scattering parameters of the Co/Cu/Co trilayer
used in the experiment of Ref.\ \onlinecite{myers,katine},
as they can be obtained through {\it ab-initio}
calculations, 
see Ref.\ \cite{stiles}.
The torque and conductance for normal
contacts with the same scattering parameters are also shown for 
comparison.
Note that 
the presence of the superconducting contact does not
affect the order of magnitude of the torques, or the
sensitivity of $g$ to $\theta$. Therefore the critical current 
necessary to switch the
layers' magnetic moments should be in the same range as for N
contacts ($I_{\rm crit} \sim 10^{9}{\rm  A/cm}^{-2}$ in the point 
contact geometry of Ref.\ \onlinecite{myers} and $\sim 
10^{7}{\rm  A/cm}^{-2}$ for the pillar geometry of Ref.\
\cite{katine}). This would allow, at least for the pillar geometry,
the use of bias voltages below the superconductor gap. 
The main difference from the case of
normal metal contacts is the period of the $\theta$-dependence
of $\vec{\tau}_a$ and $g$, which changes from $2 \pi$ to $\pi$.
The $\theta$-dependence of the torque $\vec{\tau}_b$ on the layer
adjacent to S is similar to what is found for the case of two 
normal leads. 
Although Fig.\ \ref{cocuco} is for a special choice of the parameters,
we have verified that these
conclusions do not depend on the detailed choice of scattering
parameters. 


In the limit when the conductance $g_N$ of the normal spacer connecting
$F_a$ and $F_b$ is 
small, and when the mixed 
transmission and reflection probabilities $T_{\uparrow\downarrow}$
and $R_{\uparrow\downarrow}$ are $\ll 1$, simple expressions can be 
obtained for the torque and the conductance,
\begin{eqnarray}
\frac{\partial\tau_a}{\partial I}& = &  
- \cos \theta \frac{\partial\tau_b}{\partial I} = 
\frac{g_N h}{16e\pi}
\left(\frac{1}{T_{a\uparrow}}-\frac{1}{T_{a\downarrow}}\right)
\sin 2\theta, \nonumber \\
g(\theta) & = & g(0) + \frac{e^2 g_N^3}{2h N_{\rm ch}^2} 
\left(\frac{1}{T_{a\uparrow}}-\frac{1}{T_{a\downarrow}}\right)^2
\sin^2\theta.
\end{eqnarray}
Note that, in 
this limit, $\vec \tau_a$, $\vec \tau_b$, and $g$ do not depend 
on the detailed properties of layer $b$ but only on the 
direction of its magnetic moment $\vec{m}_b$.

The predicted absence of a GMR in the usual sense [$g(\theta)$
is equal for $\theta=0$ and $\pi$] 
requires some clarification, as  GMR has been 
observed in multilayers with superconducting contacts~\cite{pratt}, 
in contrast to our predictions and those of Ref.~\onlinecite{lambert}. 
One possible reason for the observed GMR is spin relaxation between
the ferromagnetic layers and between the ferromagnetic layer and
the superconductor. Especially for multilayer systems, spin 
relaxation can not be neglected. Another cause can be that 
ferromagnets not only serve as spin filters, but  they also
cause scattering from minority electrons into majority electrons
and vice versa \cite{lambert}. 
Finally, for ballistic FNFS layers, quantum 
interference effects that are not considered here
can be important, thus providing another mechanism for
the conventional GMR \cite{ryzhanova}.
In each of these cases, the conductance of the multilayer
will be different for $\theta=0$ and $\theta=\pi$, although the
underlying reasons for this GMR are more subtle than
in the case of normal metal contacts.  


To conclude, we would like to emphasize that the 
stability of the perpendicular configuration is an intriguing
feature; for example, it allows for the
construction of a nanomagnet resonator with a current and
magnetic-field controlled frequency: when a current is passed through
the system such that the torque $\vec{\tau}_a$ keeps the moment 
$\vec{m}_a$ in the plane perpendicular to $\vec{m}_b$,
an applied
magnetic field parallel to $\vec{m}_b$ will make $\vec{m}_a$ precess 
in its plane with a frequency controled by the magnetic field.
 
We thank P.\ Chalsani, A.\ A.\ Clerk, 
E.\ B.\ Myers, and D.\ C.\ Ralph for
friendly and useful discussions. We also acknowledge the hospitality of
the Lorentz Center of Leiden University, where this  work was initiated.\vspace{-0.5cm}


\end{document}